# STATUS AND SPECIFICATIONS OF A PROJECT X FRONT-END ACCELERATOR TEST FACILITY AT FERMILAB*

J. Steimel[#], R. Webber, R. Madrak, D. Wildman, R. Pasquinelli, E. Evans-Peoples,
FNAL, Batavia, IL 60510, U.S.A.

## Abstract

This paper describes the construction and operational status of an accelerator test facility for Project X. The purpose of this facility is for Project X component development activities that benefit from beam tests and any development activities that require 325 MHz or 650 MHz RF power. It presently includes an H- beam line, a 325 MHz superconducting cavity test facility, a 325 MHz (pulsed) RF power source, and a 650 MHz (CW) RF power source. The paper also discusses some specific Project X components that will be tested in the facility.

## INTRODUCTION

Fermilab's future involves new facilities to advance the intensity frontier. In the early 2000's, the vision was a pulsed, superconducting, 8 GeV linac capable of injecting directly into the Fermilab Main Injector [1]. Prototyping the front-end of such a machine started in 2005 under a program named the High Intensity Neutrino Source (HINS). While the HINS test facility was being constructed, the concept of a new, more versatile accelerator for the intensity frontier, now called Project X, was forming. This accelerator comprises a 3 GeV CW superconducting linac with an associated experimental program, followed by a pulsed 8 GeV superconducting linac to feed the Main Injector synchrotron [2].

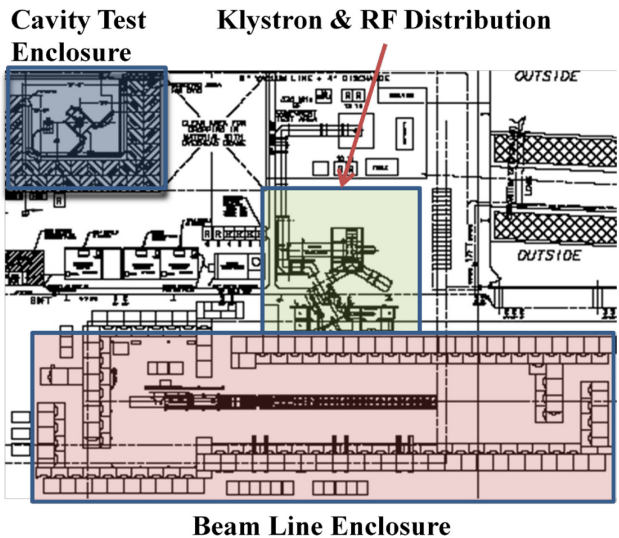

Figure 1: Layout of the Project X Front-end Test Facility

The CW Project X design is now the model for Fermilab's future intensity frontier program. Although CW operation is incompatible with the original HINS front-end design, the installation remains useful for development and testing many Project X components.

## CURRENT TEST FACILITY

The layout of the test facility is shown in Figure 1. The main components of the layout are the beam line enclosure, the cavity test enclosure, and the RF power distribution.

### Beam Line Layout and Enclosure

The current test facility beam line produces 2.5 MeV protons. These protons are generated by a 50 keV duoplasmatron ion source and then accelerated by the 2.5 MeV RFQ. A diagnostic line is located downstream of the RFQ [3]. It consists of a quadrupole triplet followed by various instrumentation devices: multiple BPMs, wire scanners, a slit scanner, a fast Faraday cup, a spectrometer magnet, and a longitudinal profile monitor. Most of the components are used to qualify the RFQ performance. The longitudinal profile monitor and fast Faraday cup are linac instrumentation experiments. Figure 2 shows the upstream end of the 2.5 MeV line.

The entire ion source and beam line are contained in an interlocked enclosure. This enclosure is designed to shield beam line operations up to 10 MeV beam energy and 2.5kW average beam power.

The 2.5 MeV beam line construction is complete; all components are connected and under vacuum. 2.5 MeV beam should be running through the diagnostic line in early April 2011.

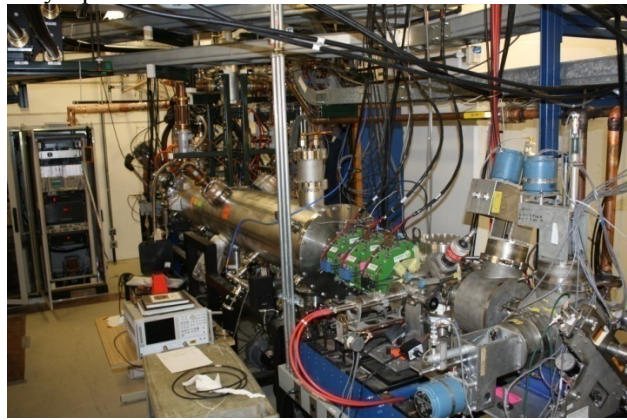

Figure 2: Front of 2.5 MeV beam line showing ion source, RFQ, quadrupoles, longitudinal profile monitor, fast Faraday cup, and wire scanner.

## 325 MHz Cavity Test Enclosure

The cavity test enclosure is used to test and condition various RF cavities outside the beam line. The enclosure is equipped with 4K liquid helium distribution, and a cryostat capable of housing a single dressed, superconducting spoke resonator. It also contains a stand and water distribution for testing room temperature spoke cavities.

A 325 MHz superconducting spoke resonator has been successfully tested both with the CW and with the pulsed high power from the klystron [4]. Work continues on testing the cavity under CW RF power conditions.

## RF Distribution

The current test facility incorporates a central 325 MHz, 2.5MW klystron RF source. This klystron is capable of pulse widths of up to 1ms with a 10 Hz rep-rate, or up to 3ms with no more than a 2.5 Hz rep-rate. Power from the klystron can be directed to the beam line components or to other test areas using a waveguide switch. Other test areas include the cavity test enclosure and a RF test cage used to qualify high power RF components that do not produce harmful X-rays.

At present, the only installed RF beam line component is a 2.5 MeV RFQ that requires up to 450 kW of power, However, splitters, couplers, and vector modulators are installed to support operation of up to six additional cavities in the beam line.

## PLANNED MODIFICATIONS

There are a number of near-term modifications planned for the test facility.

## $H^-$ Source

One of the first modifications is to replace the current proton source with an $H^-$ source. The proton ion source is capable of 25mA output current for 3ms at a 10Hz rep-rate. However, less than half of the current is due to unbound protons. The remaining fraction is due to bound states of hydrogen ($H^{2+}$ and $H^{3+}$) that are not accelerated by the RFQ. The presence of higher order species complicates measuring the ion source beam emittance and RFQ accelerating efficiency [5].

The $H^-$ source will bring the test facility more in line with the Project X plan that requires $H^-$ beam. It will greatly reduce the fraction of higher order species in the beam and will enable the use of a more diverse range of instrumentation (i.e. laser wire scanners). The source is currently being tested and qualified in a separate ion source test area. Delivery and commissioning of the new source in the Project X test facility is scheduled for June 2011.

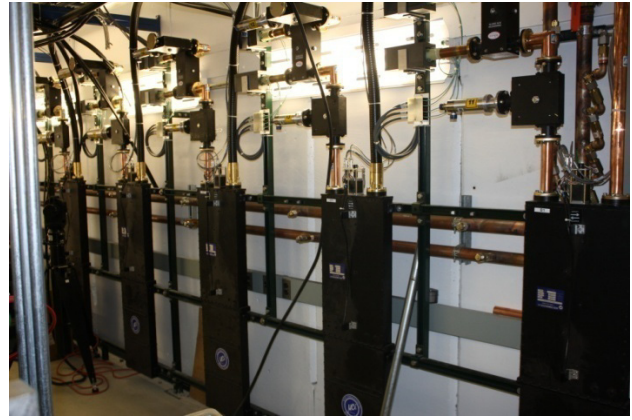

Figure 3: RF Distribution for six cavities. Large boxes are variable delay lines. Vector modulators are in front of lights.

## Six Cavity Beam Line

The HINS front-end design called for a single, high-power, pulsed 325 MHz RF source to drive multiple RF cavities in the linac. Dynamic amplitude and phase control for each RF component is performed using ferrite vector modulators [6]. Six copper cavities will be installed in the beam line downstream of the RFQ, increasing the beam energy to 3 MeV. This configuration will allow verification of the HINS multi-cavity, vector-modulator-controlled linac operational concept. Figure 3 shows the RF distribution in the enclosure, including vector modulators for the six cavities. These modulators can control up to 75 kW of RF power. A separate 550kW vector modulator is dedicated to the RFQ.

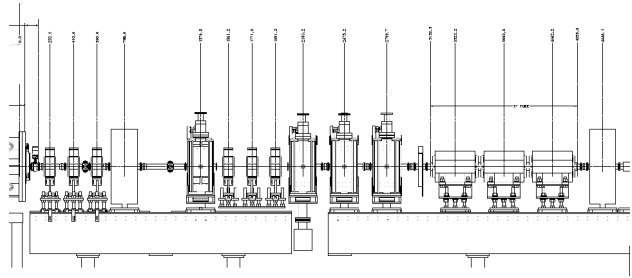

Figure 4: The acceleration section of the six cavity test. The first item after RFQ is a quad triplet, followed by a buncher cavity and spoke resonator. These are followed by a quad triplet, spoke resonator triplet, quad triplet combination. The final item before the diagnostic line is a buncher cavity.

Figure 4 shows the layout of the acceleration section of the six-cavity beam line. These cavities have all been constructed and conditioned. The required RF distribution components and utilities are installed in the beam line enclosure. The current diagnostic line will be moved downstream of the new accelerating section. A new RF control system for the klystron and vector modulators is scheduled to be completed in June 2011. This beam line installation is currently scheduled to occur in conjunction with the $H^-$ source upgrade.

### 650 MHz Cavity Test Enclosure

Plans are being made to extend the current cavity test enclosure to accommodate another test cryostat. This cryostat will support testing 650 MHz elliptical superconducting cavities for Project X. Two 30 kW CW RF amplifiers are already installed and tested. Delivery of the cryostat is not expected until 2013.

## POTENTIAL BEAM TESTS

### Linac Instrumentation

The Project X linac will benefit greatly from a non-destructive means of measuring beam profiles. The front end test facility can serve as a development and test bed for such novel linac instrumentation. Plans are already underway to test a laser wire in the diagnostic line.

### MEBT Chopper

A beam chopper is necessary to deliver proper bunch patterns to the Project X experiment [7]. This chopper must be capable of extinguishing an arbitrary pattern of single bunches from the bunch train. It will be located in the Medium Energy Beam Transport (MEBT) section of the linac, immediately downstream of the RFQ. Operation at the lowest possible beam energy reduces the activation and thermal stress on the chopper beam absorber.

Verification of successful chopper operation is an important step in proving the feasibility of Project X. The test facility described here offered the flexibility and instrumentation to perform the required chopper tests. The current test facility is capable of reaching close to 5% of the average beam current of Project X. With a variable pulse width, the facility can provide a means to safely study the thermal stresses on the chopper.

### Prototype Cryomodule

In the Project X design, the transition from normal conducting to superconducting accelerating structures occurs at the RFQ output energy. This is the lowest superconducting transition energy of any high intensity hadron accelerator in existence, and no superconducting spoke cavities have been tested with beam. It will be important to verify the operation of these cavities with beam.

Testing a cryomodule with beam will require installation of a cyrogenics distribution system in the beam enclosure. In this facility, testing will need to be done with pulsed, not CW, beam. Nevertheless, with proper RF control, studies of great value to Project X will still be possible.

## SUMMARY

Table 1 gives the summary of the front end test facility beam specifications.

Table 1: Summary of Test Beam Specifications

| | |
|---|---|
| Beam Energy | 2.5 or 3 MeV |
| Bunch Spacing | 325 MHz |
| Pulse Width/Rep Rate | Up to 1ms at 10Hz or 3ms at 2.5Hz |
| Beam Species | $H^+$ or $H^-$ |
| Ion Source Energy | 50 keV |